\documentclass[12pt,twoside]{article}
\usepackage{graphicx,cite}
\begin{document}
\parindent0cm
\renewcommand{\thefootnote}{\fnsymbol{footnote}}

\begin{center}
 {\bf \large Near Threshold $K^+K^-$ Meson-Pair Production in
  Proton-Proton Collisions\\}
\vspace{0.5cm}
 C. Quentmeier$^1$,
 H.-H. Adam$^1$,
 J. T. Balewski$^{2}$\footnote[1]{Present address: Indiana University Cyclotron Facility,
    Bloomington, Indiana 47408, USA},
 A. Budzanowski$^2$,\\
 D. Grzonka$^3$,
 L. Jarczyk$^4$,
 A. Khoukaz$^1$,
 K. Kilian$^3$,
 P. Kowina$^5$,
 N. Lang$^1$,\\
 T. Lister$^1$,
 P. Moskal$^{3,4}$,
 W. Oelert$^3$,
 R. Santo$^1$,
 G. Schepers$^{3}$,
 T. Sefzick$^3$,\\
 S. Sewerin$^{3}$
 M. Siemaszko$^5$,
 J. Smyrski$^4$,
 A. Strza{\l}kowski$^4$,
 M. Wolke$^3$,\\
 P. W\"ustner$^3$,
 W. Zipper$^5$\\
\vspace{0.5cm}
$^1$Institut f\"ur Kernphysik, Westf\"{a}lische
Wilhelms-Universit\"{a}t,\\
  D-48149 M\"unster, Germany\\
$^2$Institute of Nuclear Physics,
  PL-31-342 Cracow, Poland\\
$^3$IKP and ZEL, Forschungszentrum J\"ulich,\\
  D-52425 J\"ulich, Germany\\

$^4$Institute of Physics, Jagellonian University,
  PL-30-059 Cracow, Poland\\
$^5$Institute of Physics, University of Silesia,
PL-40-007 Katowice, Poland\\
\end{center}

\begin{abstract}
The near threshold total cross section and angular distributions of $K^+K^-$ pair
production via the reaction $pp \longrightarrow pp K^+K^-$
have been studied at an excess energy of Q = 17 MeV
using the COSY-11 facility at the cooler synchrotron COSY. The
obtained cross section as well
as an upper limit at an excess energy of Q = 3 MeV represent the first measurements
on the $K^+K^-$ production in the region of small excess energies
where production via the channel
$pp \longrightarrow pp \Phi$ $\longrightarrow pp K^+K^-$
is energetically forbidden. The possible influence of a resonant production
via intermediate scalar states $f_0$(980) and $a_0$(980) is discussed.\\
\end{abstract}


\section{Introduction}
Recently, detailed measurements on the $K^+$ meson production in
proton-proton collisions have been performed in the previously unexplored
near threshold region of the reaction channels
$pp \longrightarrow pK^+\Lambda$ and $pp \longrightarrow pK^+\Sigma^0$
\cite{bal96,bal98,sew99,bil98}.
On the other hand, there is a lack of data on the elementary $K^-$ meson production in the
proton-proton scattering, especially in the region of
low excess energies. The
reaction channel with the lowest threshold energy is given
by the associated $K^+K^-$ meson pair production via the reaction
channel $pp \longrightarrow pp\,X$, $X$ = $K^+K^-$.
Therefore, measurements on the threshold production of negatively
charged kaons (m($K^\pm$) = 493.677 MeV/c$^2$ \cite{pdb}) have
to be carried out
in the mass range of m(X) $\sim$ 1 GeV/c$^2$.
A study of this mass range is stimulated by the
continuing discussion on the nature of the scalar resonances
f$_0$(980) and a$_0$(980), which have been interpreted as
conventional $q\bar{q}$ states \cite{mor93}, $qq$-$\bar{q}\bar{q}$
states \cite{jaf77} or as $K\bar{K}$ molecules
\cite{wei90,loh90}.
In the latter case the possibility of a $K\bar{K}$ molecule interpretation
of the f$_0$(980) particle crucially depends on the strength of the
$K\bar{K}$ interaction, which can be probed in the near
threshold production of kaon-antikaon pairs \cite{kre97}.\\
Exclusive $K^-$ production data are also of special interest in the
context of sub-threshold kaon production experiments in
nucleus-nucleus interactions, which are expected to probe the
antikaon properties at high baryon density. Recent inclusive sub-threshold
measurements \cite{lau99} resulted in comparable $K^+$ and $K^-$ yields
at the same energy per nucleon below the production thresholds
for the elementary reactions $pp\rightarrow K^\pm X$.
To explain this observation, different models \cite{kap86} consider
repulsive and attractive interactions of kaons and antikaons within the nuclear medium,
respectively.
In addition, the antikaon potential in dense nuclear matter is
closely related to open questions in astrophysics
\cite{bro88,bro94}.
However, for detailed calculations on the medium effects
a precise knowledge of the elementary $K^+$ and $K^-$ cross sections
close to the production thresholds is needed.\\
We now present the first measurement of an absolute cross section
on the close to threshold $K^+K^-$
meson pair production via the reaction channel $pp \longrightarrow pp K^+K^-$
at an excess energy of Q = 17 MeV \cite{que00}, i.e. below the $\Phi$ meson
threshold.
Angular distributions of final state particles and particle subsystems are included.
In addition, one upper limit at an excess energy of Q = 3 MeV is reported (see also
\cite{kho00}).\\
The excitation function is generally expected to follow the four-body
phase space volume modified by the proton-proton interaction.
Final state interactions in the nucleon-kaon,
nucleon-antikaon or kaon-antikaon subsystems as well as
influences of the scalar resonances $f_0(980)$ and $a_0(980)$
on $K\bar{K}$ production might lead to deviations from this expectation.\\

\section{Experiment}
\begin{figure}
 \centerline{\includegraphics[width=11.0cm]{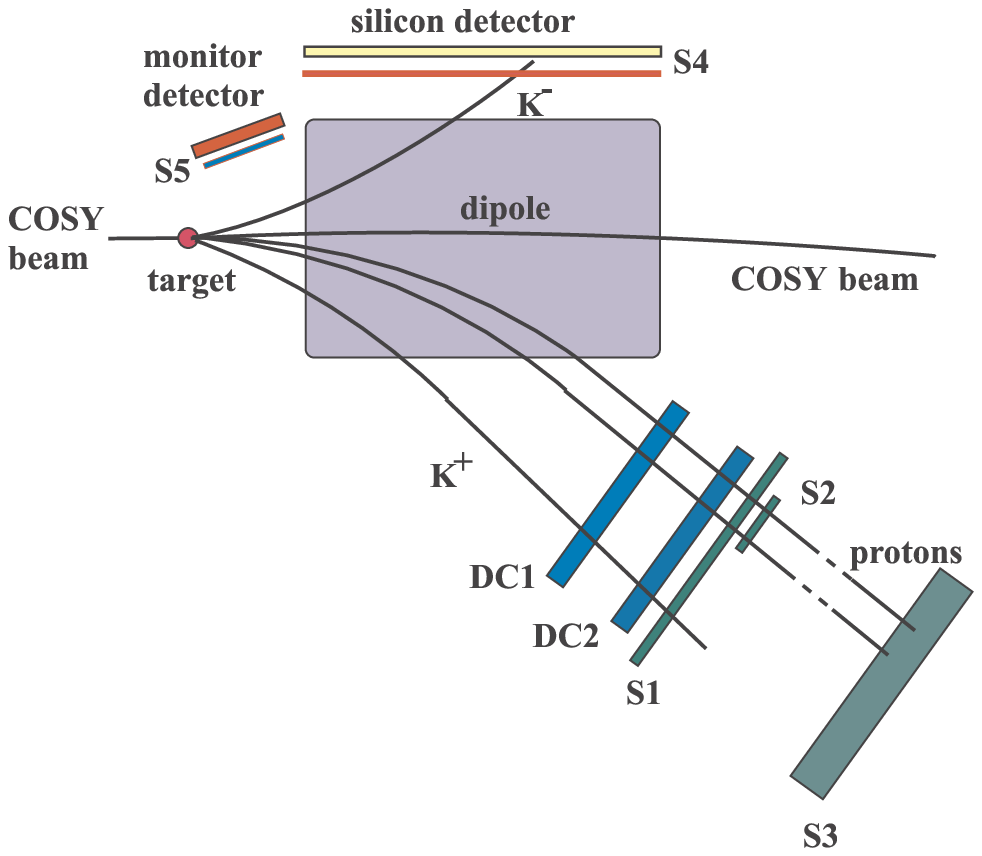}}
 \caption{Sketch of the internal beam installation COSY-11 at COSY.}
 \label{c11}
 \end{figure}

\begin{figure}
 \centerline{\includegraphics[width=11.0cm]{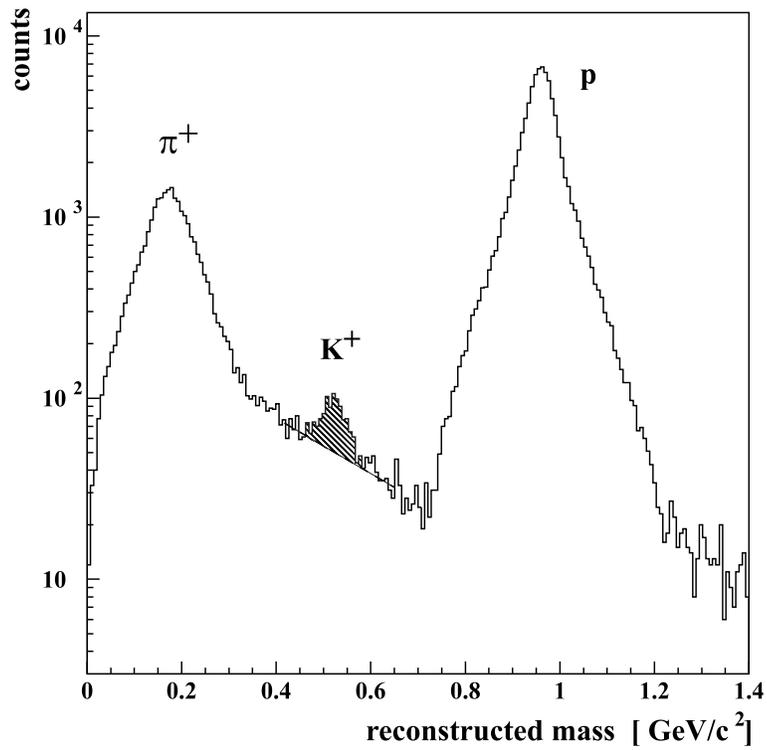}}
 \caption{Particle identification via the
 determination of the reconstructed mass for events with three reconstructed tracks.
 Shown are the reconstructed masses of all three detected ejectiles.}
 \label{invmk}
 \end{figure}

\begin{figure}
 \centerline{\includegraphics[width=13.0cm]{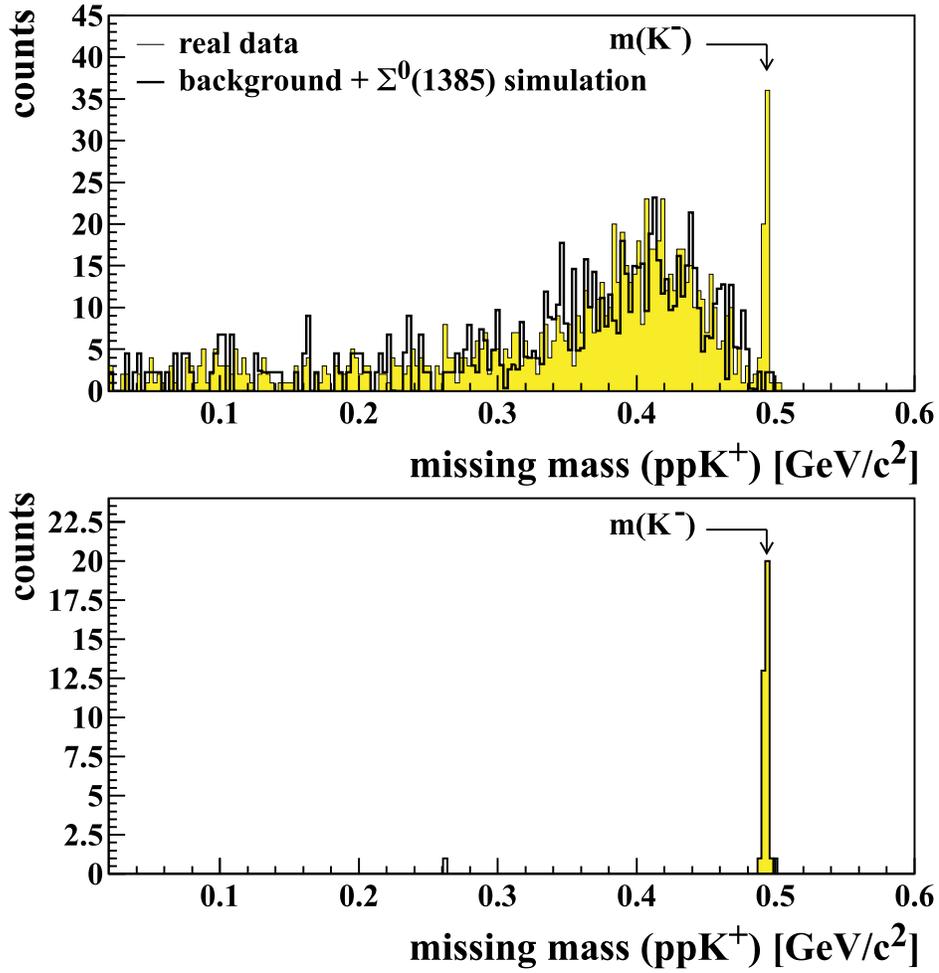}}
 \caption{Missing mass distributions of the $ppK^+$ system at an
 excess energy of 17 MeV above threshold. The picture on the top
 presents events with two identified protons and one $K^+$
 meson (thin solid line). The spectrum indicated by a thick solid line is
 a reproduction of the background distribution and is explained in
 the text.
 The lower spectrum represents events with an additional hit
 of the $K^-$ mesons in the silicon pad detector, fulfilling the $K^-$
 hit correlation of Fig.~\ref{korr}.}
 \label{mmppk}
 \end{figure}

\begin{figure}
 \centerline{\includegraphics[width=11.0cm]{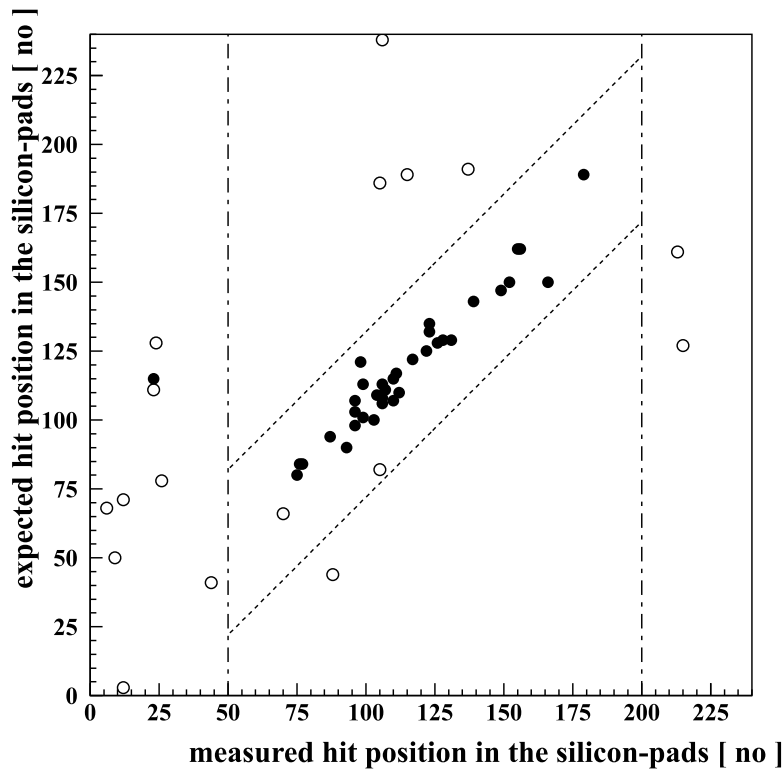}}
 \caption{
 Comparison of detected hit positions of $K^-$ mesons in the
 silicon pad detector with the calculated coordinates. Filled symbols
 correspond to events from the $K^-$-peak whereas open symbols
 represent events from the continuous background of Fig.~\ref{mmppk}
 (upper figure). The dashed and dash-dotted lines indicate cut conditions
 derived from Monte-Carlo simulations.}
 \label{korr}
 \end{figure}

Measurements on the reaction $pp \rightarrow ppK^+K^-$ have been
performed at the internal beam facility COSY-11 \cite{bra96} at
COSY-J\"ulich \cite{bec96}, using a
hydrogen cluster target \cite{dom97} in front of a C-shaped
COSY-dipole magnet, acting as a magnetic spectrometer.
Tracks of positively charged particles, detected in a set of two
drift chambers (DC1 and DC2 in Fig.~\ref{c11}), are traced back through the
magnetic field to the interaction point,
leading to a momentum determination. The velocities of these particles
are accessible by a time-of-flight path behind the drift chambers,
consisting of two
scintillation hodoscopes (start detectors S1 and S2) followed by a large
scintillation wall (S3) at a distance of $\sim$ 9.3~m, acting as a stop detector.
The overall time-of-flight resolution for events with two identified protons has been
determined to be $\sim$330 ps \cite{mos98}.
By measuring the momentum and the velocity, a particle identification
of positively charged ejectiles via the reconstructed mass determination is possible
and the four momentum vector can be completely determined.\\
The event selection for the reaction $pp \rightarrow ppK^+K^-$
was performed by accepting three-track events with two identified
protons and one particle identified as a $K^+$ meson. With respect
to the short lifetime and the opening angle of kaons in the laboratory
system, an indirect time-of-flight measurement was
used for the kaon identification. By calculating the start time of the
event in the target via the precisely determined proton trajectories
and velocities, the kaon hit in the start detectors
is used as stop signal. The quality of this method is demonstrated
in Fig.~\ref{invmk} with a clearly identified kaon peak.\\
The four-momentum determination of the positively
charged ejectiles yields a full event reconstruction for the
reaction type $pp\rightarrow ppK^+X$ and allows an
identification of the undetected X-particle system using the missing mass method.
The result of the measurements at an excess energy of Q = 17 MeV
with respect to the $K^+K^-$ threshold is shown in Fig.~\ref{mmppk}
(upper spectrum, thin line).
As expected from Monte-Carlo simulations
using the code GEANT-3 \cite{gea93}, including the phase-space event generator GENBOD,
a sharp peak at the
charged kaon mass with a missing
mass resolution of FWHM $\sim$ 2 MeV/c$^2$ is obvious,
corresponding to an unambiguous detection of events from the
$ppK^+K^-$ reaction.
The broad distribution in the region of lower missing masses
can be explained by contributions from $pp\rightarrow pp\pi^+X$
events, misidentifying pions as $K^+$ mesons.
Furthermore,
the production of heavier hyperons, in particular $\Lambda(1405)$
and $\Sigma^0(1385)$, may contribute to this background.
At the discussed beam energy these hyperons can be
produced via the reactions $pp\rightarrow pK^+\Lambda(1405)/\Sigma^\circ (1385)$.
Taking into account decay channels
with one decay proton in the final state, these events can also show the
requested signature of two protons and one $K^+$.
However, due to the decay modes of these hyperons, here the system $X$
of the reaction $pp\rightarrow ppK^+X$
consists of more than only one particle. Therefore, the missing mass
distribution of the $ppK^+$ system originating from these reaction
channels results in a broad distribution.\\
In order to reproduce the observed broad missing mass distribution,
both mentioned contributions have been considered.
For this purpose, $pp\rightarrow pp\pi^+X$
events have been analyzed, assuming for the $\pi^+$ mesons the mass of charged
kaons. Furthermore, using for simplicity only S-waves,
phase space Monte-Carlo simulations on the hyperon production,
applying the properties of the $\Sigma^0(1385)$ hyperon
(mass m = 1383.7 MeV/c$^2$, width $\Gamma$ = 36 MeV/c$^2$ \cite{pdb})
and isotropic angular distributions have been carried out.
The combination of both contributions can be seen in Fig.~\ref{mmppk} (thick
line). Obviously, these two effects, the misidentification of $pp\pi^+X$
events and the production of heavier hyperons, are sufficient to reproduce the
background distribution.
It should be emphasized
that the background contamination in the $K^-$ peak area is in any case very
small, leading finally to a total number of N = 61$^{+0}_{-5}$ accumulated
$K^+K^-$ events.\\
For a further verification of the $K^+K^-$ assignment, a silicon pad
detector, mounted inside the COSY dipole magnet, has been used to
determine the hit position of outgoing $K^-$ mesons.
Since the four-momentum vector of the $K^-$ meson is
accessible via the completely determined $ppK^+$ system,
the hit position in the silicon pad detector can be predicted.
Taking into account only events of the $K^-$ peak of Fig.~\ref{mmppk}
(upper spectrum),
the expected hit positions are compared with the
measured ones, as seen in Fig.~\ref{korr} (filled symbols).
The upright dash-dotted lines
indicate the region of the silicon pad
detector, which can be hit by $K^-$ mesons originating from the
$K^+K^-$ production at Q = 17 MeV excess energy.
Based on Monte-Carlo simulations, the dashed lines specify the
3$\sigma$ region for the deviation of the measured hit position from the calculated
one ($\sim$ $\pm$ 13.7 cm).
Obviously, except for one event all events marked by filled symbols fulfill these
requirements, giving additional and independent evidence for the
exclusive $K^-$ production.
Contrary to this, events from the broad structure of Fig.~\ref{mmppk}
show no correlation between the measured and
expected hit positions (open symbols).
Due to geometrical acceptance only a few of these events have a hit
in the pad detector at all.\\
Taking into account only $ppK^+$ events with an additional $K^-$ candidate
fulfilling the $K^-$ hit correlation (Fig.~\ref{korr}),
the primary missing mass distribution of $ppK^+$ events reduces to the
lower spectrum of Fig.~\ref{mmppk}. This procedure leads,
consistent with Monte-Carlo calculations to i) a
drastic reduction of the broad structure and ii)
to a 43\% decrease of the counting rate for the $K^-$ signal
due to acceptance and decay losses. According to these
results, the events represented by the $K^-$ peaks can be identified
as $ppK^+K^-$ events with hardly any background contamination.

\section{Results}
\begin{figure}
 \centerline{\includegraphics[width=11.0cm]{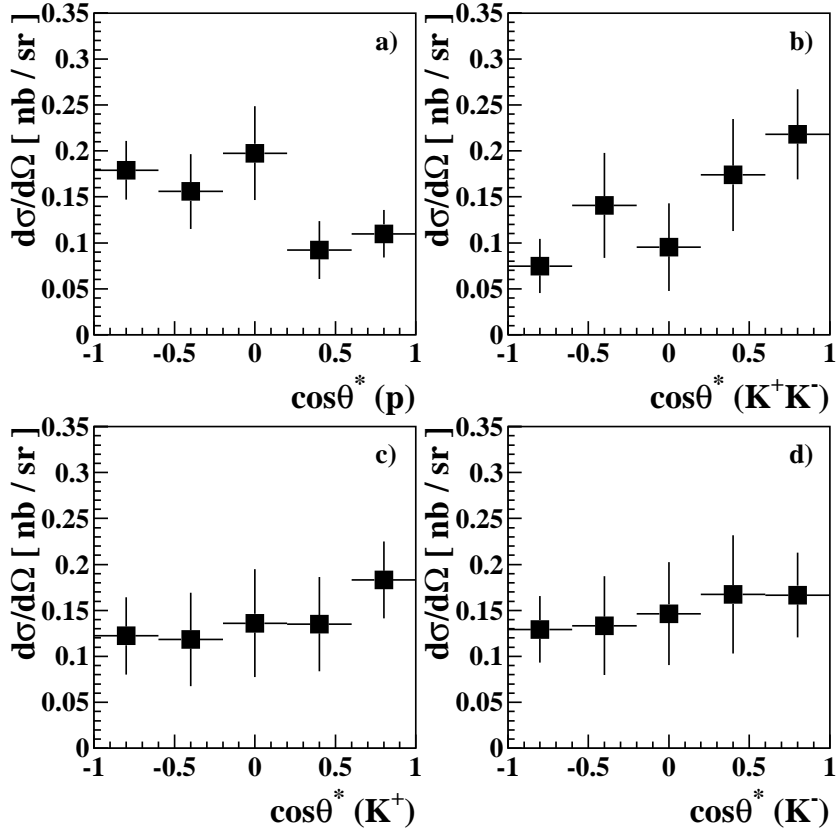}}
 \caption{
 Angular distributions in the CMS relative to the beam direction
 of the extracted $ppK^+K^-$ events for both outgoing protons (a),
 the $K^+K^-$ system (b), the $K^+$ mesons (c) and the $K^-$ mesons
 (d).}
 \label{winkel}
 \end{figure}

\begin{figure}
 \centerline{\includegraphics[width=11.0cm]{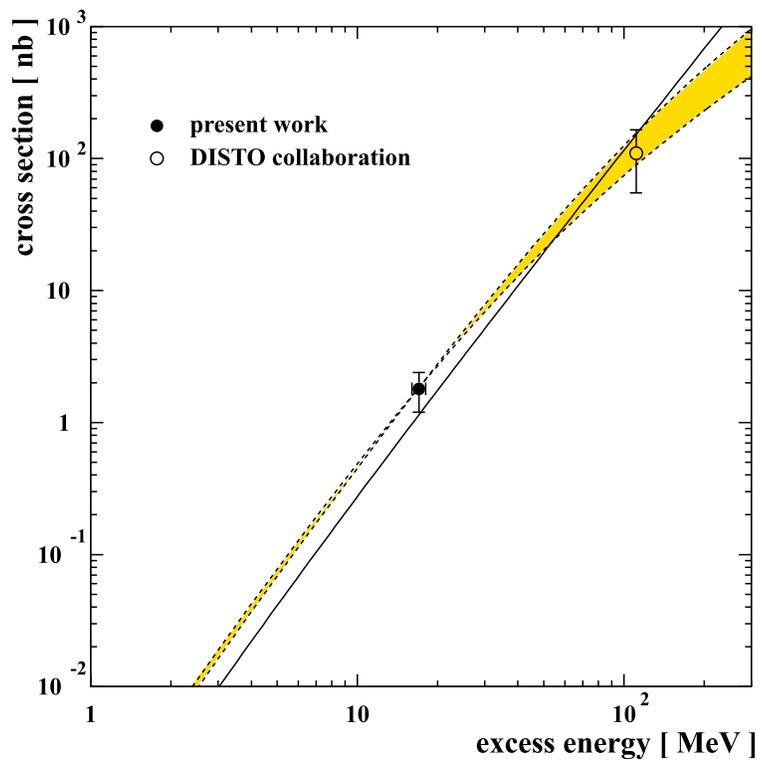}}
 \caption{
 Total cross sections for the free
 $K^+K^-$ pair production in proton-proton collisions.
 The lines are described in the text.}
 \label{kkwq}
 \end{figure}

In the following analysis we
use the events of the $K^-$ peak of the upper spectrum of Fig.~\ref{mmppk}.\\
In addition to the non-resonant production of $K^+K^-$ events, we
also consider the possibility, that the selected events
originate from other reaction channels leading to the same final state
particles. As discussed above, the production of heavy hyperons
like the $\Lambda(1405)$ and the $\Sigma^\circ (1385)$ is of minor
importance, since it appears only as a small background below a clear $K^+K^-$ signal
(see Fig.~\ref{mmppk}). Different to this it is more difficult to unravel the
contribution of the scalar resonances $f_0(980)$ and/or $a_0(980)$
from which the higher energy part of these broad resonances might
decay into $K^+K^-$ pairs \cite{pdb}.
Due to the masses of the $f_0$- and $a_0$-resonances
(m($f_0$) = 980 MeV/c$^2$ $\pm$ 10 MeV/c$^2$ and
m($a_0$) = 984.8 MeV/c$^2$ $\pm$ 1.4 MeV/c$^2$) and their
large widths ($\Gamma$($f_0$) = 40 MeV/c$^2$ to 100 MeV/c$^2$ and
$\Gamma$($a_0$) = 50 MeV/c$^2$ to 100 MeV/c$^2$) \cite{pdb},
kinematical distributions of both the resonant and the non-resonant
channels are expected to be similar.
The shape of the $pp$-missing mass distribution is sensitive to
the assumed reaction channel and therefore
offers the possibility to investigate contributions from
resonant production. However, the available statistics of the
extracted $K^+K^-$ events at Q = 17 MeV is not sufficient to
distinguish between four-body phase space and predictions on the resonant
production.
Since the structures of the $a_0$(980) and $f_0$(980) are rather similar with
respect to their effect concerning the limited acceptance of COSY-11, we
only consider the $f_0$(980) in the further discussion.\\
The accessibility of the four-momentum vectors of all
ejectiles from $ppK^+K^-$ events
allows to study angular distributions of the particles
or particle systems.
Monte-Carlo simulations on the
free $pp\rightarrow ppK^+K^-$ reaction, considering also the
$pp$ FSI and the Coulomb interaction, have been performed to
determine the acceptance of the detection system in order to
obtain acceptance corrected kinematical distributions.
The overall detection efficiencies for events from the non-resonant $K^+K^-$
production,  requiring the detection of both protons and the $K^+$
meson, were determined to be $\epsilon$(3 MeV) = $6.4\cdot 10^{-2}\,^{+43\%}_{-28\%}$
and $\epsilon$(17 MeV) = $7.4\cdot 10^{-3}\,^{+10\%}_{-13\%}$.
These quantities take into account the kaon decay,
detection and track reconstruction efficiencies
as well as the influence of the error in the absolute excess
energies, which are known with a precision of $\Delta$Q = 1
MeV, caused by the uncertainty in the determination of the absolute
COSY beam momentum ($\Delta p/p$ = $10^{-3}$).
In Fig.~\ref{winkel} angular distributions
in the center of mass system relative to the beam direction
are shown for the extracted $ppK^+K^-$ events
for both outgoing protons (a), the $K^+K^-$ system
(b), the $K^+$ mesons (c) and the $K^-$ mesons (d).
Within the statistical errors
the measured distributions of the protons and the kaons
show no significant deviation from an isotropic emission.\\
The luminosity was determined by comparing the differential
counting rates of elastically scattered protons with data
obtained by the EDDA collaboration \cite{alb97}.
The integrated luminosities were extracted to be
$\int L\,dt$ = 841 nb$^{-1}\,\pm \,1\%$\,(stat.)$\,\pm \,5\%$\,(syst.)
at Q = 3 MeV and
$\int L\,dt$ = 4.50 pb$^{-1}\,\pm \,1\%$\,(stat.)$\,\pm \,5\%$\,(syst.)
at Q = 17 MeV,
corresponding to a mean
luminosity of $L$ = 2 $\cdot 10^{30}$ cm$^{-2}$\,s$^{-1}$.\\
In Fig.~\ref{kkwq} the present result at Q = 17 MeV (filled symbol)
and a data point from the DISTO collaboration \cite{bal99},
neglecting the contribution from $\Phi$,
are plotted as function of the excess energy. These data represent
the available world data for the $K^+K^-$ production
via the reaction channel $pp\rightarrow pp K^+K^-$ in the near threshold
region (threshold: p$_{beam}$ = 3.30175 GeV/c).\\
The total cross section at an excess energy of Q = 17$\,\pm$\,1 MeV has
been determined to be
$\sigma$ = (1.80\,$\pm$\,0.27$^{+0.28}_{-0.35}$)
nb, including statistical
and systematical errors, respectively \cite{que00}.
The overall systematical error arises from uncertainties in the
determination of
the detection efficiency (8\%),
the luminosity (5\%),
the COSY beam momentum ($^{+10}_{-6}\%$)
as well as different models for the proton-proton FSI
\cite{dru97,mor68,nai77,noy71,noy72,mos00} (2\%).
From each of the four angular distributions in Fig.~\ref{winkel}
the total cross section was deduced and resulted in an additional error
contribution of $^{+5\%}_{-16\%}$ where the influence
of the few background counts in the $K^-$ peak is included.
The upper limit at Q = 3 MeV has been determined to be $\sigma$ $<$ 0.16 nb on
the basis of a confidence level of 95\%
\cite{que00}.
Additionally, Fig.~\ref{kkwq} shows parametrizations
on the $K^+K^-$ cross sections assuming different production
processes.
The solid line, representing a fit to the
data points on the basis of a four-body S-wave phase space expectations
including the proton-proton final state interaction (FSI),
describes the data points adequately within the error bars.
Therefore, the total cross section data points are consistent with a
description based on the free $ppK^+K^-$ production with no distinct
effects of higher partial waves or strong $K^+K^-$ final state
interactions.
Although not suggested by our previously discussed results,
one can calculate the cross section for the
$pp\rightarrow ppf_0(980) \rightarrow ppK^+K^-$
channel, leading to a value of
$\sigma (pp\rightarrow ppf_0\rightarrow ppK^+K^-)= 1.84 \,
\pm 0.29\, ^{+0.25}_{-0.33}$ nb.
Correspondingly, the dashed lines present three-body phase space
calculations for the $ppK^+K^-$ final
state via the excitation of the broad $f_0$
resonance including the $pp$ FSI and normalized to this
$\sigma(pp\rightarrow ppf_0 \rightarrow ppK^+K^-)$.
Here we assumed the $f_0(980)$ to be a Breit-Wigner
distribution with a mass of m = 980 MeV/c$^2$. The effect of the large uncertainty
about the width of the $f_0$(980)
resonance ($\Gamma$ = 40 MeV/c$^2$ to 100 MeV/c$^2$ \cite{pdb})
is indicated by the dashed area.
Nevertheless, within the error bars also this description is consistent with the
measured data.
Consequently, the two data points are in agreement with
both the assumption of a non-resonant as well as a resonant
production via the $f_0$, always neglecting effects of
higher partial waves.

\section{Summary}
At the COSY-11 facility the near threshold $K^+K^-$
production in the reaction channel $pp \rightarrow pp K^+ K^-$
has been studied at an excess energy of Q = 17 MeV.
Approximately
sixty $K^+K^-$ events have been extracted, leading to a total cross section of
$\sigma$ = 1.80 nb.
As a remarkable result we obtain, at same excess energies, a $K^-$
cross section differing from the $K^+$ cross section by
approximately two orders of magnitude.\\
The emission angle distributions of the ejectiles
in the center of mass system
show no significant deviation from expectations based on pure four
body phase space simulations.
The excitation function of
available close to threshold data on this reaction channel can
be described both by calculations based on the four-body phase space
including effects of the $pp$ FSI and the Coulomb interaction
as well as by calculations
based on three-body phase space considerations
via the broad $f_0$ resonance including the $pp$ FSI.

\vspace{0.5cm}

{\large \bf Acknowledgements}\\ \\
We appreciate the work provided by the COSY operating team and
thank them for the good cooperation and for delivering an
excellent proton beam. We also thank C. Hanhart for valuable discussions
on the reaction mechanism of the $K^+K^-$ production.
The research project was supported in part by the BMBF
(06MS881I), the Bilateral Cooperation between Germany and Poland
represented by the Internationales B\"uro DLR for the BMBF
(PL-N-108-95) and by the Polish State Committee for Scientific Research, and by the
FFE grants (41266606 and 41266654) from the Forschungszentrum
J\"ulich.

\clearpage

\clearpage

\clearpage

\clearpage

\clearpage

\clearpage

\end{document}